%% file: main.tex
\documentclass[sigconf,10pt]{acmart}
\renewcommand\footnotetextcopyrightpermission[1]{}
\input{src/usepackage}

\AtBeginDocument{%
  \providecommand\BibTeX{{%
    \normalfont B\kern-0.5em{\scshape i\kern-0.25em b}\kern-0.8em\TeX}}}

\setcopyright{none}

\acmConference[Woodstock '18]{Woodstock '18: ACM Symposium on Neural
  Gaze Detection}{June 03--05, 2018}{Woodstock, NY}

\begin{document} \sloppy 

\widowpenalty=0
\clubpenalty=0
\flushbottom

\input{src/commands}

\title{Recover as It is Designed to Be:\texorpdfstring{\\Recovering from Compatibility Mobile App Crashes by Reusing User Flows}{}}




\settopmatter{printfolios=true}

\input{src/authors}

\input{src/00_abstract}

\maketitle
\input{src/01_intro_v2}

\input{src/02_related_work}
\input{src/03_overview}
\input{src/03_api}
\input{src/04_compat_mode}

\input{src/08_user_study}
\input{src/09_evaluation}
\input{src/10_discussion}
\input{src/11_conclusion}

\newpage

\bibliographystyle{unsrt}
\bibliography{src/references}

\end{document}

%% file: src/usepackage.tex
\usepackage[normalem]{ulem}
\usepackage{cleveref}
\usepackage{siunitx}

\usepackage{pgfplots, pgfplotstable}
\pgfplotsset{compat=1.18}
\usepackage {gnuplottex}
\usepackage{multirow}
\usepackage{colortbl}

\crefformat{section}{\S#2#1#3} 
\crefformat{subsection}{\S#2#1#3}
\crefformat{subsubsection}{\S#2#1#3}

\usepackage{pifont}
\usepackage{subcaption}

\usepackage{listings}

\usepackage{tikz}

\usepackage[htt]{hyphenat}

\usepackage{tabularx}
\usepackage{ragged2e}
\usepackage{booktabs}
\usepackage{ltablex}
\newcolumntype{Z}{>{\RaggedRight\centering\arraybackslash\hspace{0pt}}X}
\newcolumntype{M}{>{\RaggedRight\arraybackslash\hspace{0pt}}c}
\newcolumntype{L}{>{\RaggedRight\arraybackslash\hspace{0pt}}l}

%% file: src/commands.tex
\newcommand{\proj}{\emph{RecoFlow}}
\newcommand{\libsc}{\color{red}{LibSC}}
\newcommand{\zswitch}{ZygoteSwitch}
\newcommand{\controller}{\texttt{\color{red}{compat-ctl}}}
\newcommand{\pkgtool}{\color{red}{compatibility OS packaging tool}}
\newcommand{\Pkgtool}{\color{red}{Compatibility OS packaging tool}}
\newcommand{\codegen}{SSI translation code generation tool}
\newcommand{\transformer}{\color{red}{Parcel Transformer}}
\newcommand{\pretender}{Pretender}
\newcommand{\pbdtool}{\color{red}{VizFlow}}
\newcommand{\devtool}{UFGen}
\newcommand{\userflow}{user flow}
\newcommand{\Userflow}{User flow}
\newcommand{\UserFlow}{User Flow}

\definecolor{darkgreen}{RGB}{34,139,34}
\definecolor{DARKGREEN}{RGB}{34,139,34}

\newcommand{\ready}{{\color{DARKGREEN}{[READY FOR FEEDBACK]}}}

\newcommand{\dk}[1]{#1}
\newcommand{\dkc}[1]{{\leavevmode[{\color{darkgreen}{dk: #1}}]}}

\newcommand{\sj}[1]{{\leavevmode\color{magenta}{[SJ: #1]}}}
\newcommand{\sjl}[1]{[\textcolor{magenta}{[sj: #1]}]}

\newcommand{\hjy}[1]{{\textcolor{blue}{
\it [HyungJun: #1]
}}}
\newcommand{\hj}[1]{{\leavevmode\color{blue}{#1}}}

\newcommand{\sk}[1]{\textcolor{red}{[sk: #1]}}

\newcommand{\sjhan}[1]{\textcolor{purple}{[sjhan: #1]}}

\newcommand{\cp}[1]{\textcolor{teal}{[cp: #1]}}

\newcommand{\yj}[1]{[\textcolor{cyan}{yjk: #1}]}

\newcommand{\etal}{\textit{et al.}}
\newcommand{\eg}{\textit{e.g.},}
\newcommand{\ie}{\textit{i.e.},}
\newcommand{\cf}{\textit{c.f.},}
\newcommand{\xmark}{{\ding{55}}}
\newcommand{\cmark}{{\ding{51}}}
\newcommand{\rewrite}[1]{\textcolor{blue}{#1}}
\newcommand{\eat}[1]{}
\newcommand{\about}[1]{{\textbf{#1: }}}
\newcommand{\tbd}{\textcolor{red}{\textit{?}}}

\newcommand{\tabcaption}[1]{\caption{#1} \vspace{-.15in}}

%% file: src/authors.tex
\author{Donghwi Kim}
\affiliation{%
  \institution{Samsung Electronics}
}
\email{dh.tony.kim@samsung.com}

\author{Hyungjun Yoon}
\affiliation{%
  \institution{KAIST}
}
\email{diamond264@kaist.ac.kr}

\author{Chang Min Park}
\affiliation{%
  \institution{Yahoo}
}
\email{pcm4150@gmail.com}

\author{Sujin Han}
\affiliation{%
  \institution{KAIST}
}
\email{sujinhan@kaist.ac.kr}

\author{Youngjin Kwon}
\affiliation{%
  \institution{KAIST}
}
\email{yjkwon@kaist.ac.kr}

\author{Steven Y. Ko}
\affiliation{%
  \institution{Simon Fraser University}
}
\email{steveyko@sfu.ca}

\author{Sung-Ju Lee}
\affiliation{%
  \institution{KAIST}
}
\email{profsj@kaist.ac.kr}

%% file: src/00_abstract.tex
\begin{abstract}

Android OS is severely fragmented by API updates and device vendors' OS customization,
creating a market condition where vastly different OS versions coexist. 
This gives rise to
\textit{compatibility crash} problems where Android apps crash on certain Android versions but not on others.
Although well-known, this problem is extremely challenging for app developers to overcome due to the sheer number of Android versions 
in the market that must be tested.
We present \proj{}, a framework for enabling app developers to automatically recover an app from a crash by programming user flows with our API and visual tools.
\proj{} tracks app feature usage with the user flows on user devices and recovers an app from a crash by replaying UI actions of the app feature disrupted by the crash. 
\dk{To prevent recurring compatibility crashes, \proj{} executes a previously crashed app in \textit{compatibility mode} that is enabled by our novel Android OS virtualization technique.}
Our evaluation with professional Android developers shows that our API and tools are easy to use and
effective in recovering from compatibility crashes.

\end{abstract}

%% file: src/01_intro_v2.tex
\section{Introduction}



\begin{figure*}[t]
    \centering
    \includegraphics[trim={0 8mm 0 1.2cm},clip,width=\linewidth{}]{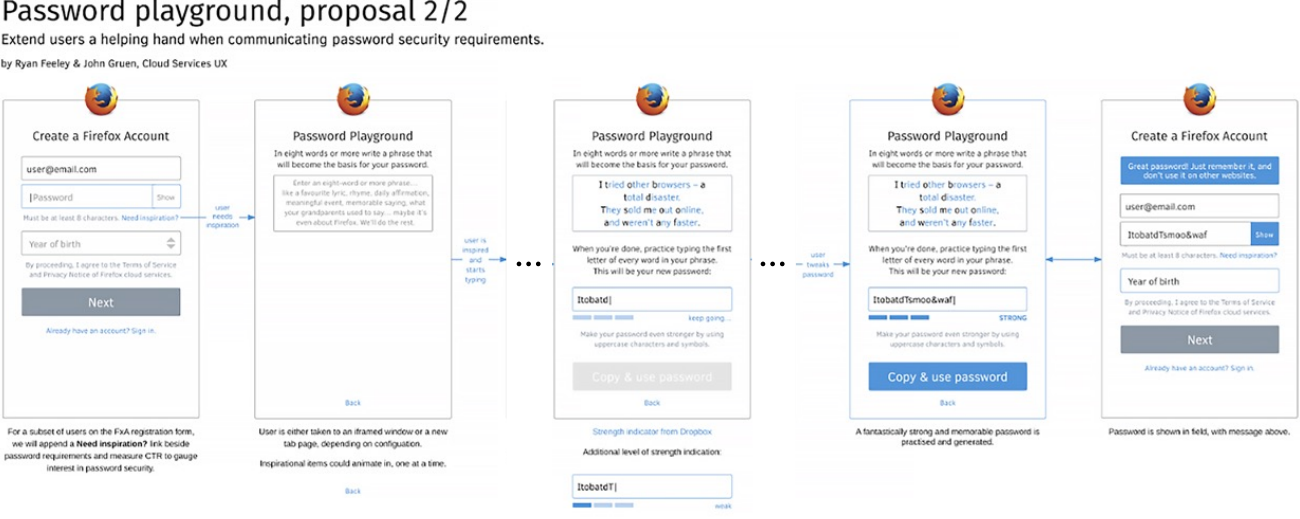}
    \Description[User flow diagram for setting a new password in the mobile Firefox app.]{User flow diagram for setting a new password in the mobile Firefox app. There are five screenshots. From left to right, each screenshot is the screen that users see when they set a new password. Each screenshot is connected with an arrow with text describing the action that users have to take to get to the next screenshot, such as "typing the new password."}
    \caption{The user flow of mobile Firefox app's new password setting feature created by Firefox app developers~\cite{MozillaC1:online}. It describes required UI actions and their order for setting a password.
    (We omitted 3 out of 8 steps for readability.)
    }
    \label{fig:firefox_userflow}
\end{figure*}



In 2022, about 950,000 Android apps were released in the market for 2.8 billion Android devices~\cite{Numberof2:online,android_users}.
Android app developers strive to create useful app features, present them effectively on the user interface (UI), and execute them stably on user devices for the huge user base.
When designing an app's UI, they optimize UI actions (\eg{} clicks) for possible user intents (\eg{} setting a password for an app account) to be easy and efficient. They often create a \textit{user flow}~\cite{Understa6:online}, a diagram of UI actions to be taken for a user intent, and improve its usability over design iterations (see figure~\ref{fig:firefox_userflow}).
When releasing the app to users, they thoroughly test the user flows on the app to ensure stable execution on user devices. They typically target a very low crash rate ($<$ 0.5\%)~\cite{CrashFre60:online, Acceptab58:online} and spend 25-35\% of the development costs for app testing~\cite{AppDevel89:online}.

\looseness=-1
Despite their efforts, developing a stable Android app is still challenging due to compatibility crashes, which are caused by incompatibility between an app and the OS version that it runs on.
Compatibility crashes are difficult to eliminate due to the open nature of Android platforms. To support 24,000 different Android device models in the market~\cite{Thereare14:online}, the Android Open Source Project (AOSP) allows device vendors to customize Android OS for a specific device. For example, a foldable phone vendor customizes Android OS to adjust UI layouts when the phone is folded, and a low-end device vendor customizes Android OS's animation quality to optimize performance. When Android's yearly major version updates are released, the device vendors customize the new versions again for their devices. During AOSP's version updates and device vendors' customizations, Android OS APIs are added, modified,
and deleted, often in a way that app developers do not expect and are hard to deal with.
Thus, apps occasionally crash when deployed on Android OSes that mismatch App APIs and OS APIs~\cite{ficfinder}.
Compatibility crashes are difficult for app developers 
to predict, reproduce, and fix because it is not practically
feasible to test their apps on all Android variations  
customized for diverse devices.


Consequently, compatibility issues are found in every Android OS version and diverse vendors' devices. 
The issues remain until app developers fix them or incompatible OS versions and devices are removed from the market. 
We investigated the top 50 open-source apps downloaded more than a million times~\cite{F-Droid,GitHub}. 
We found 99 and 57 compatibility issues due to Android version updates and vendors' OS customization, respectively. 
FicFinder~\cite{ficfinder}, Pivot~\cite{pivot}, and CiD~\cite{cid} found 191, 10, and 7 compatibility issues from 27, 10, and 7 already published 
open-source Android apps, respectively. Moreover, when we examined the source code update history of the apps that fixed their compatibility issues, the issues lasted for 30 months on average. 

Since app developers cannot prevent every compatibility crash,
Android apps should recover from their crashes.
However, developing a general recovery solution for diverse apps is problematic because (1) recovering an app's state often requires custom logics specific to app internals, and (2) existing fault tolerance methods, such as application checkpointing~\cite{CRIU3:online}, cannot guarantee the complete elimination
of recurring crashes. Although these methods can recover an
app and restart it, the root cause of the problem still exists, leading to the possibility of further crashes.

To overcome these difficulties, we propose to repurpose user flow as the basis for a general recovery solution. Our basic approach is that, when an app crashes, we restore the app by replaying previously taken UI actions for a disrupted user intent. To identify users' UI actions that belong to a user intent, we
provide app developers with a convenient API and a visual code generation tool to express user flows in their apps. For replaying, we provide a
record-and-replay system that matches a user's UI actions with a user flow and replays it. 
Using user flow is more advantageous than app state restoration because (1) it is agnostic to app internals
as it only deals with UI actions, and (2) it does not involve the restoration of the app state that
might be corrupted by compatibility issues.
In addition, we designed a new execution environment called
\emph{compatibility mode} that prevents our recovery process from encountering the same
compatibility crash, which we describe in \cref{sec:compatibility-mode}. 

To demonstrate the idea, we build a crash recovery system, \proj{}, that addresses all significant challenges via UI-driven compatibility crash recovery.
The first challenge is tracking user intents on a user device \dk{for later recovery}. Our API guides app developers to consider one user intent at a time and develop a user flow, which is a graph of possible UI actions to be matched for the intent. With our API and developer tool, the developers can visually \dk{migrate a user flow on a design document to a Java program} by selecting UI elements on app screens. During an app execution, \proj{} holds UI action records that match with unfinished user flows and replays them on a crash.


The \dk{second} challenge is in avoiding repeating the same compatibility crashes.
With \proj{}, AOSP can micro-virtualize Android OS on a user device and execute an app on a compatible guest Android OS with other apps on the host user device OS.
The key idea is (i) designating a small number of Android OS releases as \textit{compatibility OS}es,
(ii) providing an execution environment that acts as a compatibility OS to resolve OS API mismatches,
and (iii) enabling an app to run on a compatibility OS execution environment when it crashes on the default OS installed on a user device.
\proj{} automatically detects an app crash, boots a compatibility OS within a few seconds~(2.5~sec on average),
and re-launches the crashed app in the compatibility OS.

Our system evaluation demonstrates that \proj{} effectively recovers commodity open-source applications by replaying our user study participants' UI action traces. Our developer study participants developed app crash recovery logic for diverse app features with our API. The cost for \proj{}'s compatibility mode is small 
while it effectively hides compatibility crashes.
The \proj{}'s compatibility mode successfully avoids
compatibility crashes of one real-world app and five custom apps. 
When enabled, the \proj{}'s compatibility mode induces
only a marginal delay ($2.7\%$) to app execution with 38.7~MB 
additional memory.

We make the following contributions:
\begin{enumerate}
\item \proj{} minimizes mobile user experience (UX) disruption from compatibility crashes by providing an API for recovering a crashed app to the state before the last crash. App developers can write easy-to-maintain recovery code with our API. 
\item \proj{}'s recovery API addresses \textit{developer-guided} record-and-reply (R\&R) challenges of \textit{users’ UI actions}. For example, \proj{} enables developers to select only UI actions required for crash recovery in a user’s UI action history. In existing work, a developer or a user must R\&R their own UI actions.
\item \proj{} is the first compatibility mode app execution framework for Android.\footnote{We will open source \proj{} with this paper's publication.}
\item \proj{} implements a novel Android OS virtualization technique
in a real system that effectively executes an app in a virtualized OS yet incurs only minimal overhead.
\end{enumerate}

%% file: src/02_related_work.tex
\section{Related Work}


There are a few areas of research relevant to \proj{} and we survey these in this section.

\noindent\textbf{Tracking user intents using UI actions:} Predicting user intents from their interaction traces has been crucial for mobile apps and the web to improve user engagement, increase retention, and establish a monetization strategy~\cite{Howtoide92:online}. Google Analytics~\cite{Analytic43:online}, DataDog~\cite{RUMAndro44:online}, and other user tracking services allow developers to collect users' UI interaction traces in the cloud. However, unlike \proj{}'s user intent tracking API, they do not provide a model or API for mapping individual UI actions to user intention, leaving developers to implement it themselves. 

\noindent\textbf{UI action emulation:} Programmed UI action emulation has been adopted for mobile apps and the web for testing~\cite{UIApplic91:online, 8811942, kasik_toward_1996}, app prototyping~\cite{kimXDroidQuickEasy2019}, and automation~\cite{Selenium65:online, pu_semanticon_2022, leshed_coscripter_2008, li_sugilite_2017, little_koala_2007}. While a developer generates, programs, or records UI actions to be played in those cases, \proj{}'s user intent tracking API enables a developer to select users' UI actions to be replayed on user devices. ReCDroid~\cite{8811942}, GIFDroid~\cite{10.1145/3510003.3510048} and Yakusu~\cite{10.1145/3213846.3213869} are automatic Android app crash reproduction systems. They use natural language information in a crash report to quickly create a short UI action trace that reproduces a crash. Their UI action trace generation is similar to \proj{}'s selective UI action replay, as both aim to play minimal UI actions navigating an app to a previously crashed state. However, they do not preserve user-specific app context (\eg{} a specific item a user selected before a shopping app crashes) as it only cares about crash reproduction.

\begin{figure}[t]
    \centering
    \includegraphics[width=\columnwidth]{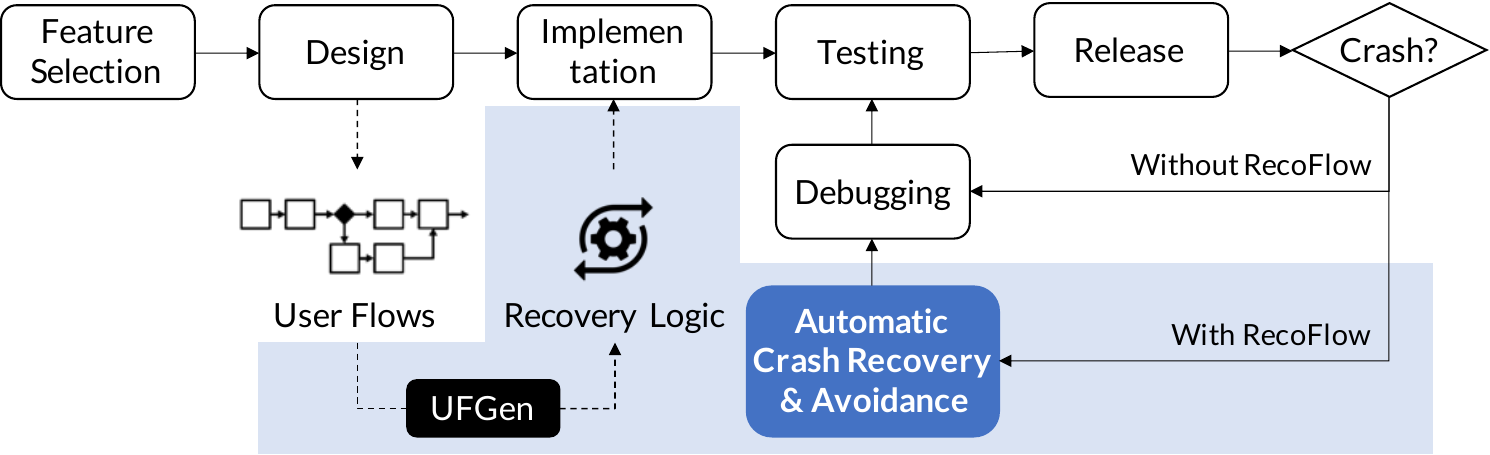}
    \caption{App development workflow with \proj{}.
    }
    \Description[Diagram of app development workflow with \proj{}.]{Diagram of app development workflow with \proj{}. There are boxes with texts (from left to right) "Feature Selection", "Design", "Implementation", Testing", and "Release." Each box represents a step in the standard app development workflow and is connected with arrows starting from the left box and ending in the right box. There are more parts that describe \proj{}'s involvement in the app development workflow. First, there is an arrow going out of the "Design" box that points to a symbol meaning User Flows. The User Flows symbol is connected to another symbol meaning Recovery Logic and the arrow has text "UFGen," which is a component of \proj{} that helps developers generate user flow code. Then, there is an arrow starting from the Recovery Logic symbol to the "Implementation" box. This portion demonstrates that recovery logic required by \proj{} is extracted from the design of the app with UFGen. Second, the rightmost box, which has the text "Release," is connected to a diamond-shaped box with the text "Crash?". There are two arrows coming out of the crash box. One arrow is connected to the "Debugging" box, which leads back to the "Testing" box. This flow describes the usual app debugging process without \proj{}. Meanwhile, the other arrow is connected to a box with the text "Automatic Crash Recovery \& Avoidance," which is connected to the "Debugging" box. This flow describes the app debugging process with \proj{}. With \proj{}, an app can provide automatic crash recovery and crash avoidance while the developer debugs the app. }
    \label{fig:developer_workflow}
\end{figure}


\noindent\textbf{Crash recovery:} Many third-party libraries assist in recovering from an Android app crash by restoring an app's state after a crash~\cite{Sunzxyon25:online,osamarad35:online}, check-pointing an app process~\cite{CRIU3:online}, repeating a failed sub-routine~\cite{Failsafe35:online}, and defining fallback for failure cases~\cite{mmin18Sa11:online,rerlangg97:online,JessYanC38:online}. However, a recovered app could crash again as the cause of the crash still remains in the recovered app. Rx~\cite{rx} recovers a request-response style server software crash. On a crash, it reconfigures the execution environment (\eg{} CPU scheduling algorithm and memory alignment configuration). It then restores the software to a previously created checkpoint and replays the last requests that incurred the crash. If the software crashes again, Rx repeats restoration and replay in a different execution environment until the crash disappears.
\proj{} is different from previous methods as it recovers an Android app in compatibility mode where the cause of the compatibility crash disappears.

\noindent\textbf{Compatibility mode of Windows OS:}
Windows OS provides a `compatibility mode’ of execution. When enabled, it links the program with old version OS libraries and APIs~\cite{windows_shims}. Although \proj{}'s compatibility mode has a similar goal, it boots a compatibility OS and launches an app with the compatibility OS while Windows OS changes libraries to be dynamically linked during an app launch. \proj{} not only replaces host OS's APIs with compatibility OS's but also executes OS API functions in compatibility OS's environment. 


%% file: src/03_overview.tex
\section{Overview of Developing an App with \proj{}}

\begin{figure}[t]

    \centering
    \includegraphics[width=\columnwidth]{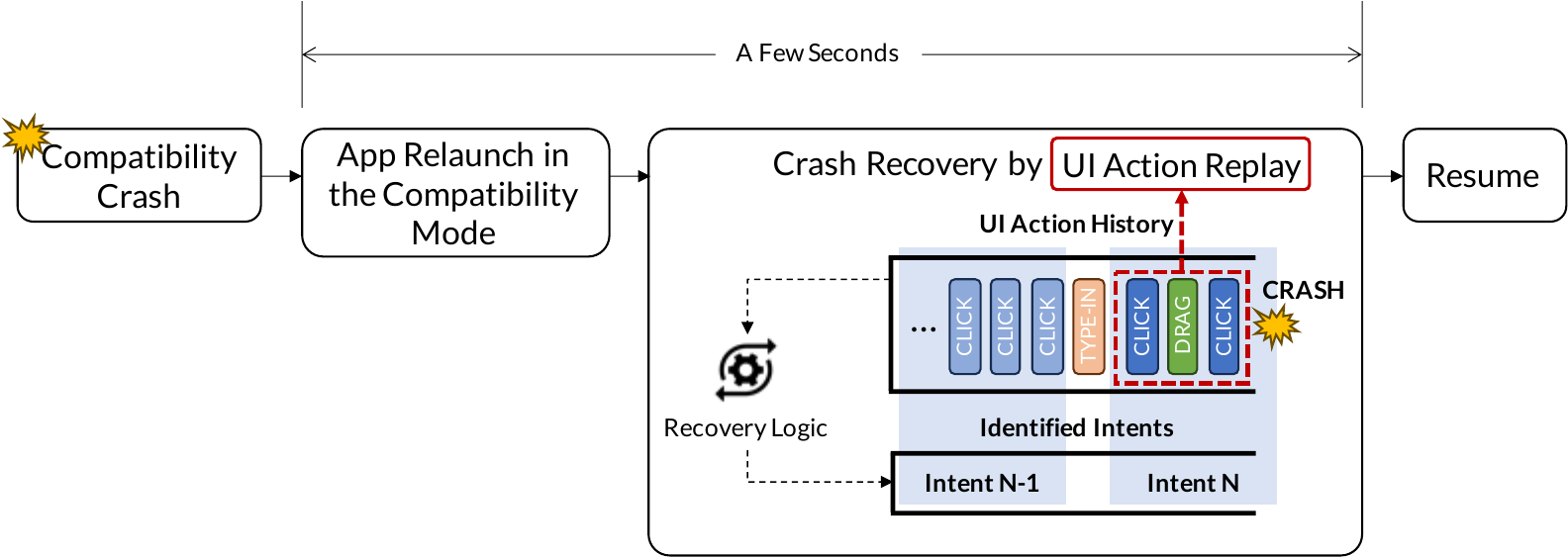}
    \caption{Automatic crash recovery with \proj{}. Recovery logic, which is a user flow written in Java, classifies UI actions into user intents, and \proj{} recovers an app by replaying UI actions of an intent disrupted by the crash.}
    \Description[Diagram of automatic crash recovery with \proj{}.]{Diagram of automatic crash recovery with \proj{}. There are four boxes with text (from left to right): "Compatibility Crash", "App Relaunch in the Compatibility Mode", "Crash Recovery by UI Action Replay", and "App Use Resume." There is a double-sided arrow on top of the "App Relaunch in the Compatibility Mode" and "Crash Recovery by UI Action Replay" boxes to indicate that those two processes take only a few seconds. The "Crash Recovery by UI Action Replay" box contains details of UI Action Replay. When a crash happens, a subset of UI actions that matches the current intent is selected based on the recovery logic. Then, only those selected UI actions are replayed. }
    \label{fig:automatic_recovery}
\end{figure}
With \proj{}, Android app developers can automatically recover their apps from a crash by reusing user flows created for app UI design. It requires two additional easy steps in the app development process: (i) integrating our library into the apps and (ii) programming existing user flows in Java with our API and visual tool.

A typical app development process involves feature identification, design, implementation, testing, release, and maintenance, as shown in figure~\ref{fig:developer_workflow}. Developers commonly create user flow diagrams during UI design for possible user intents. A user flow diagram is a graph of app screens where UI elements to interact for a user intent are highlighted. The whole graph represents the sequence of UI actions required when fulfilling the intent and the graph is improved over design iterations. When the graph is finalized, the developers implement and test it on Android devices they have. Since testing the app on all device models and OS version combinations is practically impossible, they select a few major OS versions and the most popular device models. When the app is released to users, the app could sometimes crash due to incompatibility with an untested combination of a specific OS version and device model. The crash is reported to the developers by crash reporting libraries (\eg{} Crashlytics~\cite{Firebase20:online}); however, releasing a fixed app usually takes a long time as they often do not possess the problematic device model. 


\proj{} enables the developers to run their latest apps on incompatible user devices with minimal effort while they fix compatibility issues. While implementing the app, the developers transform the user flow diagrams to Java code with our API. They can visually perform this task with \devtool{}, our user flow code generator. \devtool{} takes the APK of an app under development and mirrors the running app's screen in an Android emulator. To obtain the Java code for a user flow diagram, the developers must navigate to the app screens in the user flow and visually select highlighted UI elements on \devtool{}.
While doing so, \devtool{} automatically converts each node, edge, and highlighted UI elements to a Java object with our API and generates a corresponding code segment.
To enable \proj{} for the app, the developers must paste the generated code in the app's source directory
and initialize \proj{} during the app launch.

When an app with \proj{} runs on a user device, \proj{} records all UI inputs and matches them with the user flow codes. If the app crashes, \proj{} re-launches the app in \textit{compatibility mode}, which prevents recurring compatibility crashes. If a user flow was matched at the moment of the crash, \proj{} recovers the app state by replaying the UI action that matched the user flow. Figure~\ref{fig:automatic_recovery} describes this process. 
This way, \proj{} enables app developers to fix compatibility issues while their apps installed on user devices avoid and recover from compatibility crashes.
We further discuss our API and the compatibility mode in Sections~\ref{sec:recovery_api} and~\ref{sec:compatibility-mode}.


%% file: src/03_api.tex
\section{Visually Programming a Crash Recovery Logic with \proj{} API}
\label{sec:recovery_api}

With \proj{}, an app developer can develop a crash recovery logic for each app feature. To do so, one has to express each user flow for the features with our API. This allows a developer to effectively capture what a user must do to accomplish a goal, which can later be used to identify what needs
to be replayed to resume the work a user started before a crash.
In addition, we provide a developer tool, which we call \devtool{}, that automatically
generates user flow code that uses our API.

\subsection{User Flow in Stages}

While a user flow is typically a graph of app screens and UI actions making transitions between them, we define it as a graph of UI action sets. This enables robust user flow matching against diverse user traces. 

To understand our design for user flows, consider a scenario where a user sends a message on a
chatting app. There are infinitely many possibilities for UI action sequences to accomplish the
task---a user may type a message and hit the send button right away, or a user may start typing a
few characters but then scroll the chat history for some time and later come back to finish the message,
etc. This means we cannot use an exact sequence of UI actions as a user flow since we
would need infinitely many user flows to capture a single scenario.

Thus, we define a user flow more abstractly as a directed graph, where vertices are called
\textit{stages} and edges represent stage transitions. There are two types of stages for a user
flow. The first type is the initial stages that mark the beginning of a user flow. The second type is
intermediate stages that represent in-between UI actions that a user can perform after starting a
user flow but before finishing it.
Each stage is represented as a \textit{set} of one or more UI actions, and a stage transition from
stage \textit{A} to a connected stage {B} occurs if the current stage is stage \textit{A} and \emph{one} of
the UI actions in stage \textit{B} occurs. If a UI action from stage \textit{A} keeps
occurring, no stage transitions occur.
Otherwise, if a UI action is neither in stage \textit{A} nor in any stages connected from \textit{A}
occurs, the user flow terminates.


In a stage, a UI action is defined by a UI element where the UI action occurs and a UI action type, \eg{} touches and typings. We use VPath, our extension of XPath, to specify a UI element on the screen. For example, ``\texttt{//view[@class="android.widget.Button" and @text="Click Me"]}'' matches with a ``Click Me'' button. A VPath must always be matched with a single UI element to select the exact UI element to play an action in replay-time. During a UI action replay, \proj{} plays a UI action on a UI element that matches with its VPath, and it aborts the recovery if the VPath matches with 0, 2, or more UI elements. 

\subsection{Example Scenario}
\label{sec:example_scenario}

Suppose a chat app developer wants to develop crash recovery logic for a \textit{poll creation}
feature of her app. The feature allows users to create a poll in a group chat room. Using
\proj{} API, she takes five steps to develop the crash recovery logic.


\noindent\textbf{Step 1, structuring a user intent with stages:} To track a user intent, \proj{} API requires her to express the intent in stages.
She plans to create a user flow with two stages: the starting-poll stage which a user clicks the ``Poll'' button in a chat room, and the composing-poll stage which the user fills in the poll title and poll options.


\begin{figure}
    \centering
    \hfill
    \begin{subfigure}[t]{.49\linewidth}
         \centering
         \includegraphics[width=0.60\textwidth]{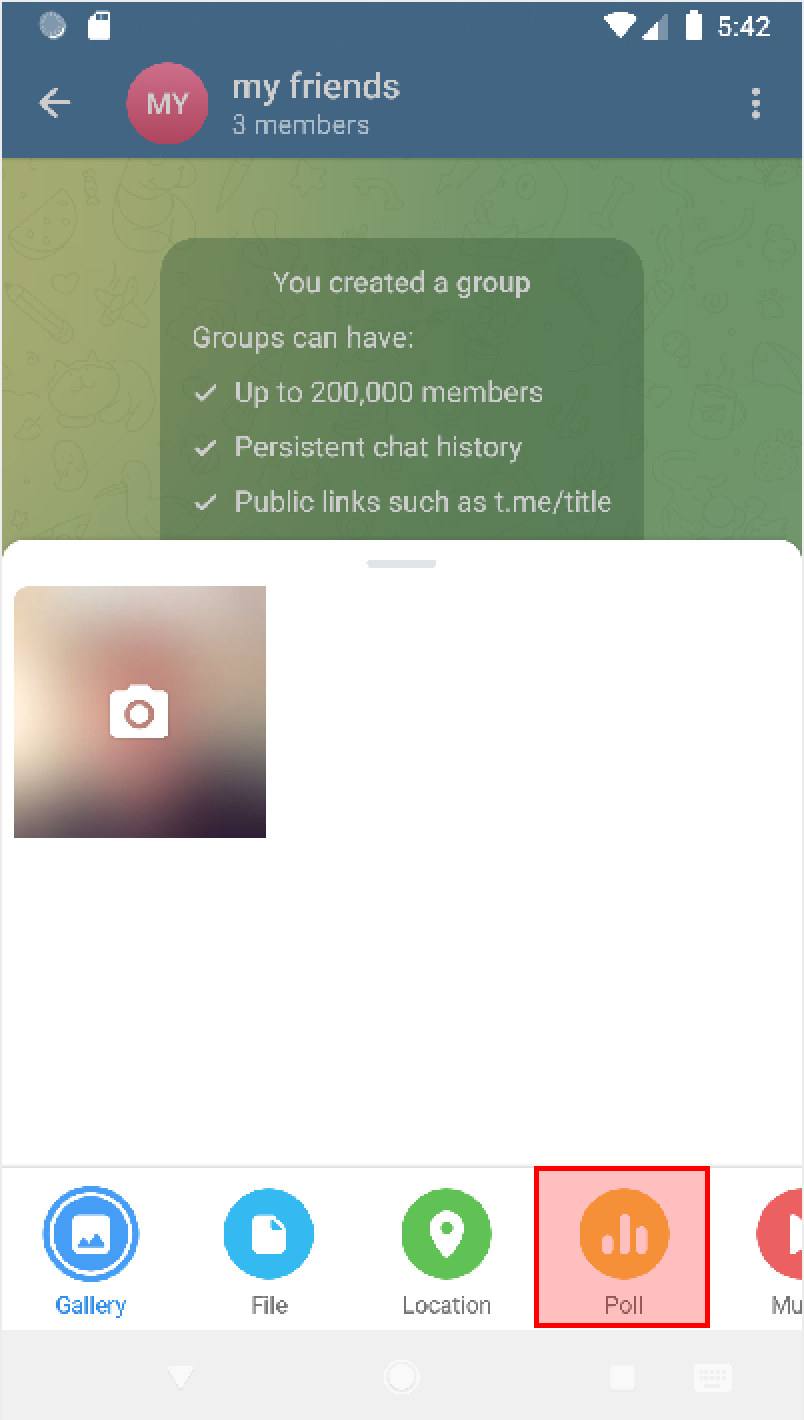}
         \Description[A screenshot of the Telegram app where the ``Poll'' button is highlighted in red.]{A screenshot of the Telegram app where the ``Poll'' button is highlighted in red.}
         \caption{Starting-poll stage.}
         \label{fig:starting_poll_stage}
     \end{subfigure}
     \hfill
     \begin{subfigure}[t]{.49\linewidth}
         \centering
         \includegraphics[width=0.60\textwidth]{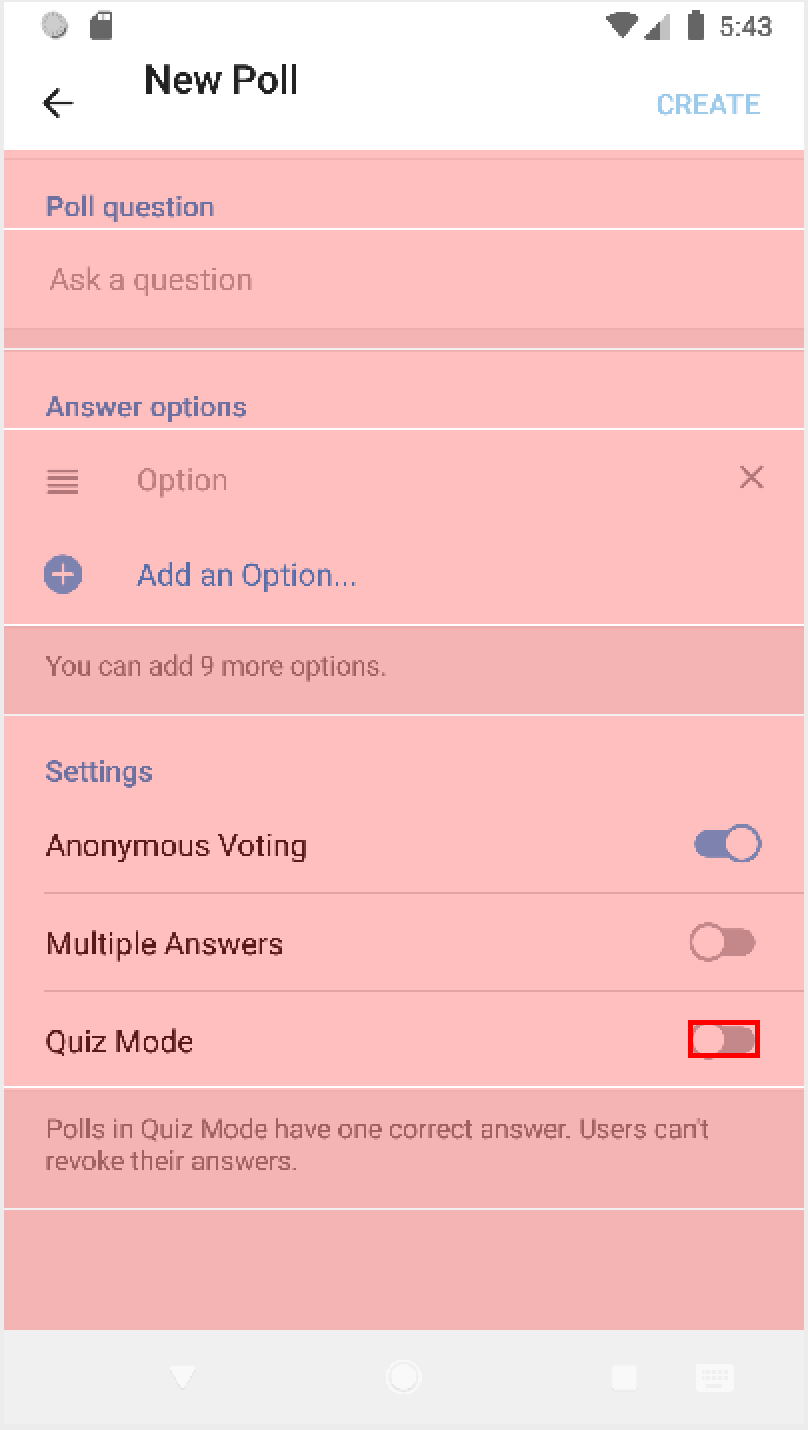}
         \Description[A screenshot of the Telegram app for composing a poll.]{A screenshot of the Telegram app for composing a poll. Every field, including the poll title, the poll options, and other poll configuration fields, is highlighted in red.}
         \caption{Composing-poll stage.}
         \label{fig:composing_poll_stage}
     \end{subfigure}
     \hfill
    \caption{The corresponding UI elements of UI actions included in the two stages in~\cref{sec:example_scenario} are filled with red.
    }
    \label{fig:stage_examples}
\end{figure}

\noindent\textbf{Step 2, visually programming each stage with \devtool{}:} She opens an IDE and launches her app in an Android emulator. She then opens \devtool{} to program each stage visually.
She wants to include a single UI action of clicking the ``Poll'' button to the starting poll stage because a user must do so to create a poll. Therefore, she clicks the ``Poll'' button on the app screen mirrored on the left side of \devtool{} (see \cref{fig:starting_poll_stage}). When she clicks the ``Generate Stage'' button on \devtool{}, it creates Java code creating a new \texttt{Stage} class instance with a single \texttt{UIActionFilter} class instance describing the ``Poll'' button click. She copies the code to the clipboard and starts creating the composing poll stage. Creating the composing poll state is similar to creating the starting poll stage except that she selects all UI elements on the poll creation pane by dragging the mouse cursor on them (see \cref{fig:composing_poll_stage}). Before generating a stage code, she unselects the ``Create'' button on the poll creation pane because clicking it ends the intent; hence, it is not part of a poll creation intent.



\noindent\textbf{Step 3, defining a user flow with the stages:}
She creates a custom user flow class in the IDE by extending \texttt{Userflow} class in \proj{} API. In her user flow class, she must implement an abstract method, \texttt{startingStage()}, which returns the starting stage of her user flow. She pastes previously generated \texttt{Stage} instance creation code in the method and connects the two stages by calling \texttt{addNextStage()} method. Finally, she makes her \texttt{startingStage()} method to return the starting poll stage instance.

\noindent\textbf{Step 4, preparing for a UI action replay:}
\proj{} provides \texttt{Userflow.prepareReplay()} callback which is called before starting the recovery. In this example, she wants to navigate the re-launched app to the chat room because her user flow starts by clicking the ``Poll'' button in the chat room. To do so, she implements \texttt{prepareReplay()} callback to navigate the app to the chat room before the replay begins. She calls \texttt{prependUIActionsToReplay()} method in the callback to prepend UI actions for clicking a proper chat room before the recorded UI actions replay. 

\noindent\textbf{Step 5, initializing \proj{}:} To enable crash recovery, she calls \proj{}'s \texttt{initialize()} method during her app's initialization and passes her user flow as a parameter.

\noindent\textbf{Intent tracking and recovery on user devices:} When her app runs on a user device, \proj{} compares every UI action on the app with \texttt{UIActionFilter} in her user flow's starting stage; \ie{} clicking the ``Poll'' button. When a user clicks the button, \proj{} records that action and transits the user flow to the starting poll stage. When the user interacts with any UI elements on the poll creation pane except the ``Create'' button, the user flow transits to the composing poll stage and continues recording UI actions as long as the subsequent UI actions match. If a crash occurs at this moment, \proj{} will restart the app in compatibility mode and recover the app by replaying recorded UI actions. If a user escapes the composing poll stage by clicking the ``Create'' button, clicking the back button, or performing any other UI actions not included in the stage, recorded UI actions and the user flow's tracking state are cleared. 

\subsection{Design Features}

We design \proj{} API carefully to cover diverse use cases and handle errors for app developers.

\label{sec:recording_with_vpath}

\noindent\textbf{Recording a UI action with VPath:}
Existing UI action recording methods fall into one of these two types: recording x-y coordinates on a screen or recording UI element and UI action type (\eg{} click). Neither of them meets our need---flexible enough to record diverse UI actions (\eg{} pinch and free-form drawing on a painting app) and descriptive enough to embed semantic information (\eg{} clicking a button named ``Poll'') at the same time. Therefore, when a UI action matches a user flow, we record both VPath matching with a UI element where the action occurs and the relative x-y coordinates of raw UI events on the UI element. When replaying the UI action, we find a UI element matching with the VPath and then replay a raw UI action after converting recorded relative coordinates into absolute screen coordinates by using the found element's position on the screen.

\noindent\textbf{Error handling:}
To place UI action replay always under a developer's control, we replay a UI action whose VPath matches with only one UI element on the screen. We abort the replay after timeout (2 seconds by default) if the VPath matches with no UI element, and we abort the replay immediately if it matches with more than one UI element. According to our developer study~(\cref{sec:developer_study}) and user study~(\cref{sec:user_study}), we could not find any cases that \proj{} replays UI actions out of target user intents (\eg{} due to a developer's mistake). 

%% file: src/04_compat_mode.tex
\section{\proj{} Compatibility Mode}
\label{sec:compatibility-mode}

We provide brief background about Android OS (\cref{sec:background}), explain \proj{}'s compatibility mode (\cref{sec:compat_mode_overview}),
and the required AOSP's support for enabling \proj{} on billions of existing Android devices (\cref{sec:ecosystem}).


\subsection{Android OS Background}
\label{sec:background}

\proj{} virtualizes only a part of the Android OS (the app framework) for its compatibility mode. The virtualized app framework resides in an additional Zygote, a special system process we inject for compatibility mode. We provide the background for understanding what the app framework is and why and how \proj{} virtualizes the app framework.

\noindent\textbf{The Android app framework:} The app framework is a set of Java and C++ libraries that provide essential
run-time services either directly by itself or indirectly by communicating with other system services. For example,
the app framework provides the Android OS API and the Dalvik Java virtual machine. The app framework is always linked
to an Android app; without it, an app cannot function properly. \proj{} virtualizes this app framework
to provide an execution environment that behaves as a compatibility OS.

\noindent\textbf{System services and their RPC interface (SSI):}
When an app accesses a system resource (\eg{} a display), it makes a request to the app framework via the Android OS API.
The app framework then communicates with an appropriate system service to handle the request.
Android has various system services, such as the Window Manager that decides which app to show on the display and the
Activity Manager that decides when to start, pause, and stop an app. These system services exist as user-space processes and 
define an RPC interface that the app framework uses to access their functionality. We call this RPC interface between the app
framework and the system services the \textit{System Service Interface (SSI)}. Different Android OS versions might have different SSI
definitions due to OS version updates or vendor customization.

\looseness=-1
\noindent\textbf{Booting the Android OS and launching an app with Zygote:}
The Android OS is based on Linux and its booting process is similar to that of Linux. It uses \texttt{init.rc} as its
boot script that performs various initialization tasks such as setting up necessary environment variables, mounting disk
partitions, etc. A notable difference between Android and Linux is the use of a special process called Zygote, a system daemon that is the parent of every Android app process. It is the process that handles app launches---it receives a request to start an app and forks a new process that loads and executes the app. Since Zygote is
linked to the app framework and every app process is a fork of Zygote, all app processes are automatically linked to the
app framework. We create an additional Zygote for \proj{} to boot a compatibility OS.


\subsection{\proj{}'s Compatibility Mode}
\label{sec:compat_mode_overview}



 The goal of \proj{}'s compatibility mode is eliminating Android OS API mismatches between an app and an Android OS by placing a compatibility OS layer. 
This goal translates into four technical objectives (O1-O4).
First, the compatibility mode should execute an app in a compatibility OS's app execution environment~(O1).
Second, an app's local data files (\eg{} a social-media app's log-in tokens and a photo app's photos)
should persist over the compatibility mode~(O2).
Third, the compatibility mode should be transparent 
to the host OS and other apps as if a compatibility mode app
is a regular app~(O3).
For example, a regular Android OS on a user device should start, pause, and stop a compatibility mode app as it does for regular Android apps. Another example is that a compatibility mode chat app should launch a regular PDF viewer app just as a regular chat app would when a user clicks a PDF file link. 
Fourth, the compatibility mode should be lightweight to run on mobile devices~(O4).

\begin{figure}[t]
    \centering
    \hfill
    \begin{subfigure}[t]{.49\columnwidth{}}
        \centering
        \includegraphics[width=\columnwidth{}]{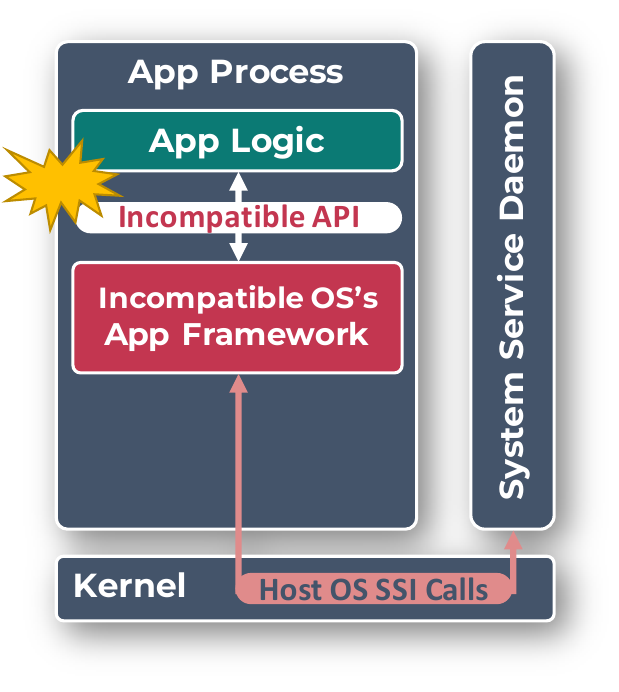}

        \Description[An Android app architecture diagram having a compatibility issue.]{An Android app architecture diagram having a compatibility issue. It has three boxes representing an App process, system service daemon, and Kernel. In the app process box, there are two smaller boxes, one representing app logic and the other representing app framework. They are connected by a double-sided arrow with text saying ``Incompatible API''. The app framework box and system service daemon box are connected by a double-sided arrow saying ``Host OS SSI calls'' which goes through the Kernel box.}
        \caption{API mismatches in the Android OS layer.}
        \label{fig:api_snippet}
    \end{subfigure}
    \hfill
    \begin{subfigure}[t]{.49\columnwidth{}}
        \centering
        \includegraphics[width=\columnwidth{}]{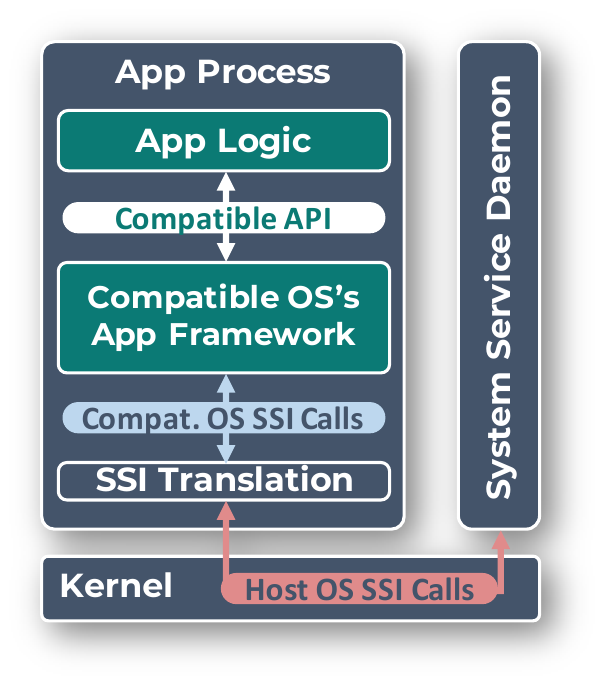}

        \Description[An Android app architecture diagram having no compatibility issue.]{An Android app architecture diagram having no compatibility issue. It has three boxes representing an App process, system service daemon, and Kernel. In the app process box, there are three smaller boxes, one representing app logic, another representing a compatible OS's app framework, and the other representing SSI translation. The app logic box and the compatible OS's app framework box are connected by a double-sided arrow with text saying ``Compatible API''. The SSI translation box is between the compatible OS's app framework box and the system service daemon box, and it connects them with double-sided arrows. The arrow between it and the compatible OS's app framework box says ``Compatible OS's SSI calls'', and the other arrow between it and the system service daemon box says ``Host OS SSI calls''.}
        \caption{No API mismatch with \proj{}.}
        \label{fig:keepassdroid_screen}
    \end{subfigure}
    \hfill

    \caption{\proj{}'s compatibility mode app execution. \proj{} effectively removes API mismatch by executing an app with compatible OS.} 
    \label{fig:architecture}
    \hfill
    
\end{figure}

Enabling our compatibility mode essentially creates a compatibility OS's app execution environment that meets the above four objectives on a user device.
Figure~\ref{fig:architecture} describes our compatibility mode app execution.

\looseness=-1
For O1, \proj{} virtualizes a compatibility OS's components for the app execution environment, specifically, the app framework, Zygote, and the init script.
\dk{
When enabled on a user device, \proj{} adds an additional Zygote that has a compatible OS's app framework. \proj{} modifies the init script to launch the additional Zygote on a device boot. If an app crash is recorded on Android logcat, \proj{} hijacks later launch requests for the app and redirects them to the additional Zygote. By doing so, the app is launched in the compatible OS's app execution environment.
}
For O2, the host OS's app-specific data directories are shared with the compatibility OS.
For O3, \proj{} executes a compatibility mode app with the host OS's system services. Therefore, the host OS can start and stop the app and relay communication between the app and other regular apps as it does for regular Android apps.
Executing the compatibility OS's app framework and the host OS's system services together needs caution as they may have inconsistent definitions of the SSI (System Service Interface) for making RPC calls to each other. Therefore, \proj{} translates every inconsistent RPC call between them. 
Regarding O4, \proj{} is lightweight as it does not add additional program components to execute except SSI translation. In our experiments, SSI translation adds only marginal run-time overhead (see \cref{sec:microbenchmark}).

\begin{table}[t]
    \centering
    \Description[A table comparing \proj{}'s virtualization technique with HW emulation, full virtualization, and containerization.]{A table comparing \proj{}'s virtualization technique with HW emulation, full virtualization, and containerization. Different virtualization techniques are in different rows, and there are four columns, O1, O2, O3, and O4. The \proj{}'s micro-virtualization row has check marks for every column, the HW emulation and the full virtualization rows have check marks only on O1 column, and the containerization row has check marks on O1 and O4 columns.}
    \tabcaption{\proj{} and other virtualization techniques' technical feasibility for the compatibility mode.}
    \scalebox{.9}{
    \begin{tabular}{|l|c|c|c|c|c|c|}
        \hline
        \textbf{Virtualization Techniques} & \textbf{O1} & \textbf{O2} & \textbf{O3} & \textbf{O4} \\ \hline \hline
        \proj{}'s micro-virtualization & \cmark{} & \cmark{} & \cmark{} & \cmark{} \\ \hline
        Hardware Emulation & \cmark{} & & & \\ \hline
        Full Virtualization & \cmark{} & & & \\ \hline
        Containerization & \cmark{} & & & \cmark{} \\ \hline
    \end{tabular}
    } 
    \label{tab:virt_compare}
\end{table}

To our knowledge, \proj{}'s micro-virtualization is the only virtualization technique achieving the four objectives when used for a compatibility mode. Emulation (\eg{} QEMU~\cite{QEMU52:online}), full virtualization (\eg{} KVM~\cite{KVM73:online}), and containerization (\eg{} \texttt{chroot}~\cite{chrootin90:online}) have been demonstrated on Android OS. However, neither their guest OS apps share data files with host OS apps, nor their guest OS apps integrate into the host OS environment. Hence, they fail to achieve O2 and O3, respectively. Furthermore, emulation and full virtualization incur significant overhead on mobile devices. Therefore, they do not achieve O4. Table~\ref{tab:virt_compare} summarizes the comparison. 
\subsection{How to Enable \proj{}}
\label{sec:ecosystem}

To enable the \proj{}'s compatibility mode, we provide a compatibility OS packaging tool, compatibility mode patch tool, and SSI translation tool to AOSP to generate compatibility OSes, patched Host OSes, and SSI translation packs, respectively.
Using the tools, AOSP performs three tasks.

\noindent\textbf{Compatibility OS image creation:}
AOSP should make the selected compatibility OSes' images publicly available. One of those images must be downloaded to a device to enable our compatibility mode.
The tasks might involve a small amount of manual effort to modify the boot script around 300 LoC.

\noindent\textbf{System Service Interface translation:}
A compatibility OS uses the host OS's system
services. This allows the compatibility OS to be lightweight but requires SSI
translation when there are any changes in SSI due to the version difference between the compatibility OS
and the host OS. Thus, when AOSP releases a new Android OS revision, it should create an SSI translation pack
for SSI changes between the new and the previous revisions.
\dk{
While it is mandatory, SSI translation is mostly mechanical and automatable. For instance, we have translated 61 out of 5,824 SSI calls for running Android 9.0 compatibility OS on Android 8.1 host OS, and only two of them require manually writing 359 lines of translation code.
}

Note that a device vendor's OS customization might change SSI. However, their SSI changes
rarely affect \proj{}'s compatibility mode as a vendor's customization mainly 
involves \emph{adding} new services to support device-specific features, e.g., foldable screen. Since a
compatibility OS is based on AOSP that does not target any specific device, there is no need for a
compatibility OS to handle those SSI changes.
\dk{
Our analysis of 12 commercial Android device ROMs from five vendors (Samsung, Google, Xiaomi, Sony, and Oppo) across three Android OS versions (8, 9, and 10) reveals that the vendors added 1,080 SSI calls for a ROM on average while modifying 5.25 and deleting 0.16 calls. We could automatically translate SSIs for all ROMs except two Samsung Galaxy S8 ROMs having a deleted SSI call.
}

\noindent\textbf{Patching user device OS:}
To launch an app in our compatibility mode,
\proj{} needs small changes to the existing Android OS 
to boot a compatibility OS and launch an app as necessary. 
To facilitate applying this change,
we developed a compatibility mode patch tool. AOSP could use this tool to patch
existing user devices as an over-the-air (OTA) update. It only requires rebooting a device once and
does not require installing a new OS.

\subsection{Motivations for AOSP on Using \proj{}}
We expect AOSP to have the motivation to perform their tasks.
AOSP has put significant efforts into alleviating the compatibility crash problem. Despite their efforts, app developers still must fix the compatibility issues that persist in numerous OS versions.
\proj{} 
automatically detects a compatibility crash and transparently executes the prepared compatibility OS. The cost 
for the compatibility mode is small (2.7\% slowdown of apps' execution time) with only 38.7 MB extra memory. In addition, the \proj{}'s toolset automates most of what the AOSP should provide. AOSP could easily apply our patch to legacy devices via an over-the-air update.

%% file: src/08_user_study.tex
\section{Evaluation}

We implemented \proj{} with 24,515 lines of Python/Java/C/C++ code. 
We evaluate \proj{} API's (i) robustness to diverse app users with a user study, 
(ii) ease-of-use with a developer study, and (iii) ease-of-maintenance with \userflow{} maintenance study. We also evaluate \proj{}'s (iv) compatibility mode's effectiveness with a compatibility crash case study and (v) performance with a microbenchmark. 

\subsection{User Study}
\label{sec:user_study}

\begin{table}[t]
\centering
\Description[A table describing user study tasks and corresponding our user flows' stage graphs.]{A table describing user study tasks and corresponding our user flows' stage graphs. The table has four columns: app, recovery target intent, user flow, and user task. There are eight recovery target intents, ``edit a bookmark'' for the Firefox app, ``log-in'', ``search for an email'', and ``classify an email'' for the K-9 Mail app, and ``search for a chat room'', ``create a poll'', ``create a group chat'', and ``update profile'' for the Telegram app. The next column describes a user flow developed by the authors for each recovery target intent. For example, the Telegram app's ``Search for a chat room'' intent has a user flow with two stages, the ``clicking the search button'' stage and the ``typing a search query'' stage connected by an arrow. The user task column lists user tasks used for the user study. For example, a user task given for the ``search for a chat room'' intent is ``search for study group chat room and say hello in there''. Each recovery target intent has one user task except the K-9 Mail app. The following common user task for the K-9 Mail app's three recovery target tasks is given: ``Log-in to the app with a given e-mail address and password. Search for emails related to rice cake and classify them into urgent, rice cake, and reference folders.''}
\tabcaption{User flows and user tasks for \cref{sec:user_study} and \cref{sec:developer_study}.}

\scalebox{0.8}{
\begin{tabular}{|c|m{2.2cm}|m{6cm}|}
\hline
\textbf{App} & \textbf{Target Intent} & \textbf{User Task} \\ \hline
\hline

\rotatebox{0}{Firefox}
& Edit a bookmark & Search for fever remedy's effects, side-effects, and history and bookmark them in new folders accordingly. \\ \hline

\multirow{3}{*}{\rotatebox{0}{K-9 Mail}}
& Log-in & \multirow{3}{6cm}{Log in to the app with a given e-mail address and password. Search for emails related to ``rice cake'' and classify them into ``urgent'', ``rice cake'', and ``reference'' folders.} \\ \cline{2-2}
& Search for an email & \\ \cline{2-2}
& Classify an email & \\ \hline

\multirow{4}{*}{\rotatebox{0}{Telegram}}
& Search for a chat room & Search for ``study group'' chat room and say hello in there. \\ \cline{2-3}
& Create a poll & Creating a poll in the ``my friends'' chat room for surveying dinner attendees. \\ \cline{2-3}
& Create a group chat & Create a new group named ``my group''. \\ \cline{2-3}
& Update profile & Update bio to ``nice to meet you''. \\ \hline

\end{tabular}
} 
\label{tab:userstudy_tasks}
\end{table}


To evaluate whether we can develop robust recovery logic with \proj{} API, we created nine different \userflow{}s for three popular open-source apps, Firefox browser (100M+ downloads), K-9 mail (5M+ downloads), and Telegram messenger (1B+ downloads). Table~\ref{tab:userstudy_tasks} lists our \userflow{}s, their target user intents, and graphs of stages we implement for them.
To evaluate our \userflow{}s with diverse user traces, we recruited eight participants (2 females and 6 males, aged 20$\sim$27) who have used Android smartphones for at least a month. We prepared three Android smartphones, Samsung Galaxy S20, Samsung Galaxy S22 Ultra, and Google Pixel 2 XL, and installed the three apps on each phone. Before starting the user study, we gave them one of the phones and asked them to explore the three apps for 10 minutes to help them be familiar with the apps.
They were compensated with \$8 for 30$\sim$40 minutes of a user study session. Our study is IRB-approved.


\subsubsection{Online Crash Study}

For evaluating the four \userflow{}s in Firefox and K-9 Mail, we asked the participants to perform a task for each app (see Table~\ref{tab:userstudy_tasks} for task details). To collect diverse user traces, we did not provide instructions to users about which UI elements they should use for accomplishing given tasks or the specific sequence in which to use them. While they used the apps for the tasks, we randomly crashed the apps when they performed between 5 to 60 UI actions to simulate compatibility crashes. The app is automatically recovered after a crash if a participant's UI actions match with our \userflow{}. If not, we instructed the participant to open the app and continue the task. During the experiments, we recorded the smartphone screen while visualizing touch inputs for later analysis.

We manually labeled all crash cases in the screen recordings. Firefox and K-9 apps crashed 74 times between all participants during the experiments. Among 41 Firefox crashes, 10 occurred while adding a bookmark. Among 33 K-9 Mail crashes, four occurred while a participant logged in, six while searching for an email, and one while classifying an email. For all aforementioned cases except three, \proj{} correctly recovered the app to the state before the crash. For the three cases, the recovery did not start as we accidentally omitted UI actions creating new folders from the `filling in bookmark info.' stage in the Firefox app's \userflow{}. We note that this mistake affected only recovery initiation, and it is less likely to happen in practical use cases where app developers have user flows used for the app's UI design. We found participants taking uninstructed actions (\eg{} writing an email and opening a settings window) and repeating similar actions (\eg{} adding bookmarks multiple times). However, those uninstructed actions and previously completed user intents were never replayed; this suggests that our API effectively replays only the necessary UI actions as intended.

\subsubsection{Offline Crash Study}
\label{sec:offline_crash_study}

For evaluating the four \userflow{}s in Telegram, we asked the participants to perform four tasks matching with each \userflow{}. This time, we did not randomly crash the app but instead recorded every UI action a participant performed for each task in a replayable format.

After collecting the UI actions, we replayed each trace on a recovery-enabled Telegram app and simulated a crash during a replay. For a thorough evaluation, we picked one of the participants and one of the task traces of the participant. We repeated it multiple times while moving the crash simulation moment from the beginning of a trace to the end of the trace by a 1-second interval. During this experiment, we captured an app screenshot before injecting a crash and captured another screenshot when the recovery was completed. We then manually compared every pair of screenshots and classified them in a confusion matrix. We did this for every participant's task. We found 647 true positive recovery cases, 408 true negative recovery cases (\ie{} not recovered because a participant was not in any of target intents, browsing a chat room list, for example), and no false positive or false negative recovery cases. This suggests that we could develop robust crash recovery \userflow{}s with \proj{} API for Telegram. 

\subsection{Developer Study}
\label{sec:developer_study}

To evaluate how easy \proj{} API is to use, we recruited six professional app developers (one female and five males, aged 22$\sim$33).
As freelancer Android app developers, P1, P2, and P3 have worked on one, three, and one Android app development projects.
P3, P4, P5, and P6 have worked as full-time Android app developers for six months, 14 months, one year, and four years, respectively.
At the beginning of an experiment session, we explained the purpose of the experiment and \proj{} API for 50 minutes. We then asked them to develop a \userflow{} for each recovery target intent we defined for Telegram (Table~\ref{tab:userstudy_tasks}). Since we could not obtain the original user flow diagrams that Telegram app developers created, we asked the participants to develop user flows with their own understanding of the intents. We did not force the order of the tasks, but we recommended first working on a chat room search intent, the easiest one, then moving to poll creation, group chat creation, and profile update tasks. For the experiment, we loaded the Telegram app on Android Studio and modified its build script to load \proj{} API. We also wrote a piece of code (5 LoC) for calling \proj{}'s initialization function that must be called at the app launch because navigating a large Telegram source code (3.7M LoC) to find the proper place for calling it may be difficult for the developers. Lastly, we provided a \userflow{} for recovering crashes while a user shares her geolocation as a live example of our API in a chat room. They spent 2.5$\sim$3 hours for \userflow{} development, followed by 10 minutes interview. They were compensated with \$75 for the four hours of developer study. Our study is IRB-approved.
All developers except P1 attended the same experiment session but they did not discuss with one another during the experiment.

All developers except P6 could develop \userflow{}s for chat room search intent and poll creation intent. While others did not, P6 had difficulty understanding how \proj{} works. P1 also developed a \userflow{} for the profile update intent, and P2 wrote \userflow{} code for the group chat creation intent and the profile update intent; however, P2 did not have enough time to test those two intents by himself.

\begin{figure}
    \centering
    \hfill
    \begin{subfigure}[t]{.49\linewidth}
         \centering
         \includegraphics[width=\textwidth]{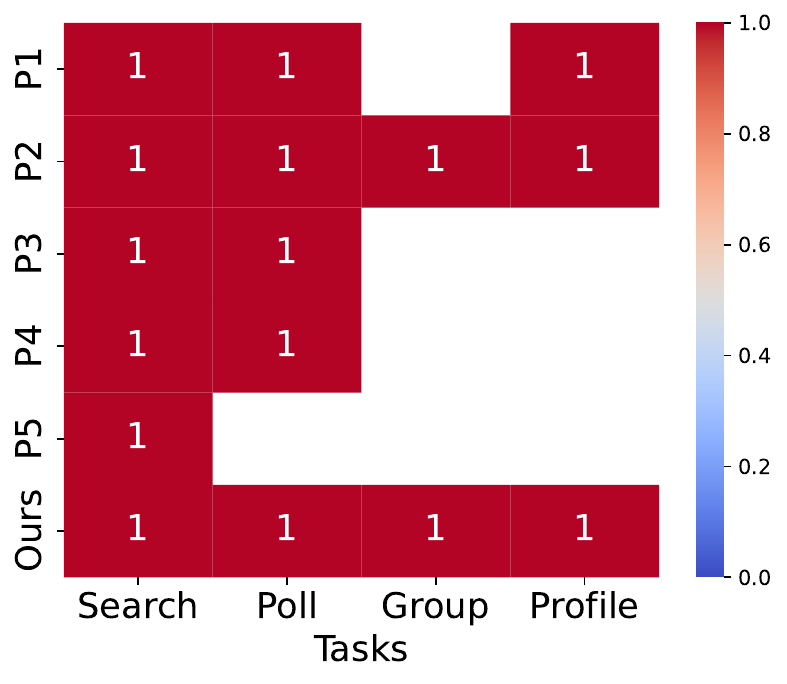}
         \Description[A heatmap reporting the recovery precision of developer study participants' user flows.]{A heatmap reporting the recovery precision of developer study participants' user flows when they are tested with synthetic UI action traces. Five developer study participants' and our user flows are in the rows, and the four Telegram user intents, search, poll, group, and profile, are in the columns.}
         \caption{Recovery precision.}
         \label{fig:dev_study_simple_precision}
     \end{subfigure}
     \hfill
     \begin{subfigure}[t]{.49\linewidth}
         \centering
         \includegraphics[width=\textwidth]{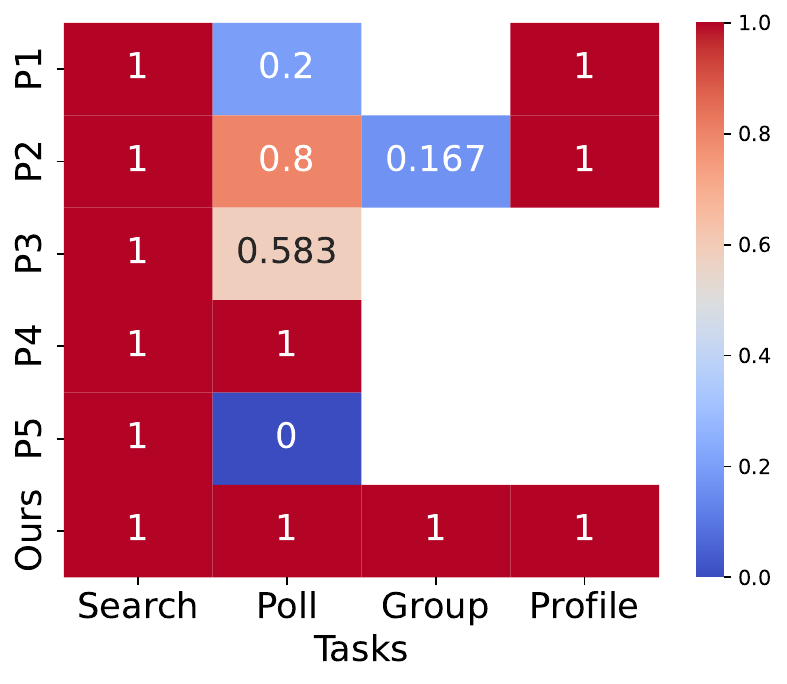}
         \Description[A heatmap reporting the recovery recall of developer study participants' user flows.]{A heatmap reporting the recovery recall of developer study participants' user flows when they are tested with synthetic UI action traces. Five developer study participants' and our user flows are in the rows, and the four Telegram user intents, search, poll, group, and profile, are in the columns.}
         \caption{Recovery recall.}
         \label{fig:dev_study_simple_recall}
     \end{subfigure}
     \hfill
    \caption{Offline crash study with synthetic UI action traces for the user flows obtained from the developer study and ours.}
    \label{fig:dev_study_simple}
\end{figure}

\begin{figure}
    \centering
    \hfill
    \begin{subfigure}[t]{.49\linewidth}
         \centering
         \includegraphics[width=\textwidth]{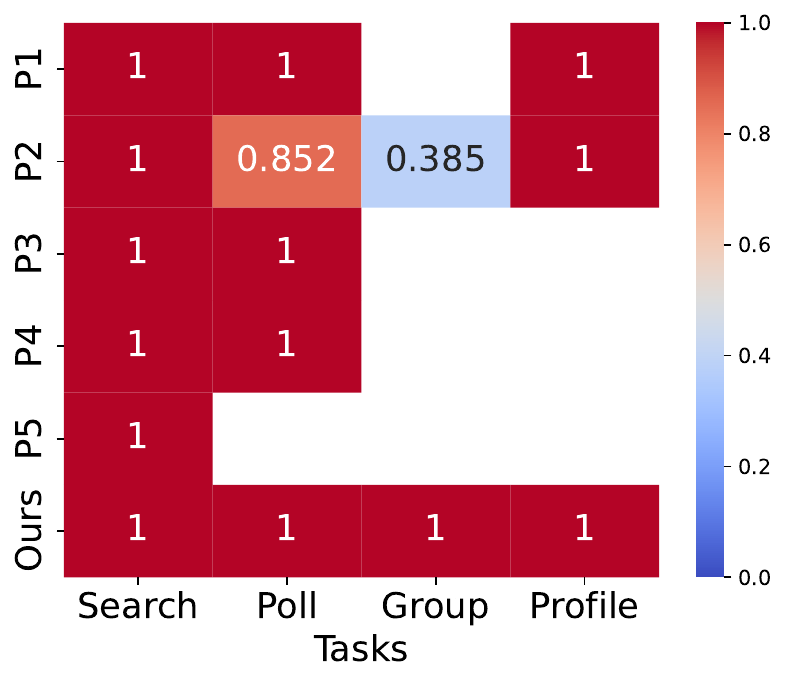}
         \Description[A heatmap reporting the recovery precision of developer study participants' user flows.]{A heatmap reporting the recovery precision of developer study participants' user flows when they are tested with UI action traces obtained from the user study. Five developer study participants' and our user flows are in the rows, and the four Telegram user intents, search, poll, group, and profile, are in the columns.}
         \caption{Recovery precision.}
         \label{fig:dev_study_real_precision}
     \end{subfigure}
     \hfill
     \begin{subfigure}[t]{.49\linewidth}
         \centering
         \includegraphics[width=\textwidth]{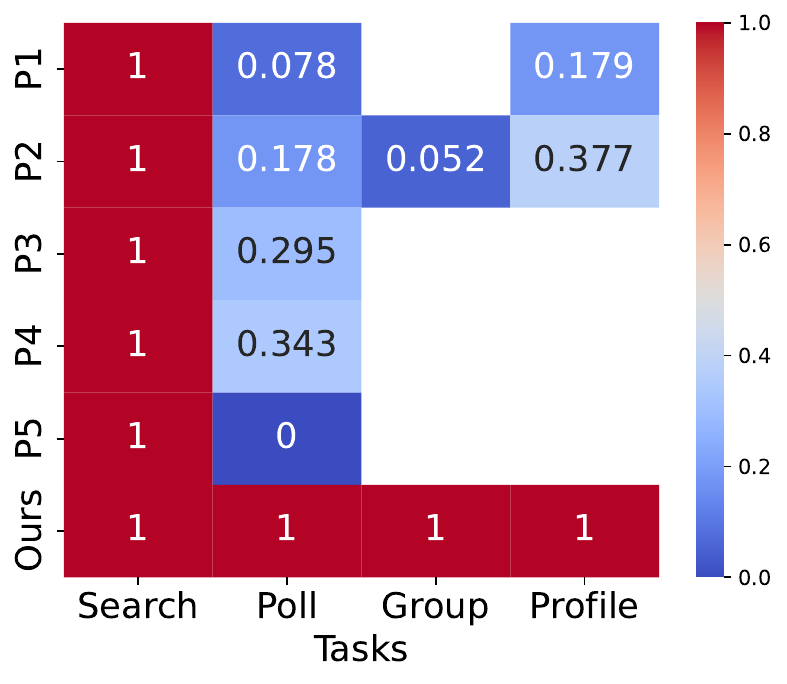}
         \Description[A heatmap reporting the recovery recall of developer study participants' user flows.]{A heatmap reporting the recovery recall of developer study participants' user flows when they are tested with UI action traces obtained from the user study. Five developer study participants' and our user flows are in the rows, and the four Telegram user intents, search, poll, group, and profile, are in the columns.}
         \caption{Recovery recall.}
         \label{fig:dev_study_real_recall}
     \end{subfigure}
     \hfill
    \caption{Offline crash study with real UI action traces for the user flows obtained from the developer study and ours.}
    \label{fig:dev_study_real}
\end{figure}

We evaluated the P1 - P5's \userflow{}s with four synthetic UI action traces that perform each Telegram user task in table~\ref{tab:userstudy_tasks}. We recorded the synthetic UI action traces while making only the UI actions essential for performing the task. With the traces, we performed an offline crash study (\cref{sec:offline_crash_study}) for the developers' user flows. Figure~\ref{fig:dev_study_simple} summarizes the results. For all developers and all user tasks, recovery precisions were 1, meaning there are no falsely triggered recoveries nor incompletely finished recoveries. For the chat room search task and the profile update task, the recovery recalls were 1, too; this means that the developers' user flows effectively recover all crash cases that occurred while a user was in the target intent. Some developers' user flows' recalls were lower than 1 for other tasks. This is mainly because the user flows do not thoroughly include all UI actions that may happen while accomplishing the given task. For example, P3's poll creation user flow stops tracking when users type in a poll option after creating one because the newly created poll option field is not included in the user flow.

We further evaluated P1 - P5's user flows with the UI action traces from user study participants. Although developers did not have enough time to get familiar with the Telegram app and \proj{} API, their user flows successfully recovered all crashes during the chat room search task. One difference from the previous evaluation using synthetic UI action traces is that the precisions of P2's poll creation user flow and group creation user flow are lower than 1. P2's UI action record of filling in a poll option matches multiple poll option UI elements when replayed because P2's VPath was not descriptive enough to distinguish them. As a result, P2's crash recovery stops when a user creates multiple poll options and fills them in. P2's group creation user flow experienced a similar issue. We note that even for P2's user flows, UI actions unrelated to a disrupted user intent were never played during the recovery. Another difference from the previous evaluation using synthetic UI action traces is that the recall values for the poll creation, group creation, and profile update tasks became lower. This is because users performed more diverse UI actions that were not included in their user flows but were still relevant to the tasks, for example, dragging an empty part of the poll creation pane to scroll the pane.





In the semi-structured interviews after user flow developments, we asked about the usability of \proj{} API and specific difficulties they experienced. P2 said that the API is comfortable once he gets used to it, and P3 and P5 noted that the API is easy and intuitive to use. P4 and P6 said the API was complicated and required more explanations. When we asked about specific difficulties they experienced, they responded that (1) selecting the exact view with \devtool{} is challenging when multiple views are overlapped on the z-axis (P2, P3, P5), (2) \devtool{} generates invalid VPath using ephemeral text for an editable text field (P1), (3) they were unfamiliar with the Telegram app (P1), (4) non-English text input was ignored (P3), and (5) they could not understand the \devtool{}-generated code (P6). We addressed (1) by selecting only the views at the top when a user visually selects views on \devtool{}, (2) by ignoring text values when a view is EditText class, (4) by ensuring Unicode compatibility, and (5) by updating our API document. We believe (3) is not a problem in practical scenarios where app developers use \proj{} on their own apps.

In the interviews, P3 and P5 expressed interest in using \proj{} API. P3 said, ``I think this API is useful for quality assurance (QA). … My boss would like to use this for QA, for users.'' P5 said, ``This will definitely help. Even if its coverage is limited, I would adopt this if such a thing exists. … From the users' perspective, they care only about UI/UX. So I am okay as it improves UX.''

\subsection{\UserFlow{} Maintenability}


A \userflow{}'s robustness against app updates is crucial for minimizing app developers' crash recovery logic maintenance burden. To evaluate that, we have tested the correct execution of our \userflow{}s for the user study (\cref{sec:user_study}) with older versions of Telegram, Firefox, and K-9 Mail. For Firefox and K-9 Mail, the oldest versions we could compile and test were 489 days and 367 days ago, respectively. We could not test versions older than 59 days ago for Telegram because the Telegram server refused to serve those old clients. In those periods, apps were actively updated; there were 16 Telegram releases, 131 Firefox releases, and 32 K-9 Mail releases. We manually tested our \userflow{}s by stimulating crashes with the oldest and a few more versions. 
For all test versions of the three apps, our \userflow{}s successfully track user intents and recover from crashes. This suggests that app developers do not have to update \userflow{}s as frequently to keep up with app updates.

%% file: src/09_evaluation.tex



\subsection{\proj{}'s Compatibility Mode Case Study}
\label{sec:case_study}

\newcommand{\compatos}{\textit{C}}
\newcommand{\incompatos}{\textit{I}}

\newcolumntype{P}[1]{>{\centering\arraybackslash}p{#1}}
\begin{table*}[]
    \centering
    \tabcaption{Execution results of apps and libraries on incompatible OSes (\incompatos{}), compatible OSes (\compatos{}), and compatibility mode of \compatos{} on host OS \incompatos{} (\proj{}).}
    \scalebox{0.80}{
    \begin{tabular}{|p{2.5mm}|p{27mm}|p{128mm}|P{35mm}|P{12mm}|}
        \hline
        & \textbf{App/Library} & \textbf{Description} & \textbf{Execution Environment} & \textbf{Results} \\ \hline \hline
        
        \multirow{9}{2.5mm}{\rotatebox[origin=c]{90}{Commodity App/Library}} &
        \multirow{3}{27mm}{\centering French Calendar App~\cite{FrenchCalendar}} &
        \multirow{3}{128mm}{Calling \texttt{setLocalOnly()} to control notifications' target devices. The method should not be called in ICS OS as ICS does not have the method, otherwise, \texttt{NoSuchMethodError} would occur.} &
        ICS (4.0) for emul. (\incompatos{}) & Error \\ \cline{4-5}
        & & & JB (4.1) for emul. (\compatos{}) & Ok \\ \cline{4-5}
        & & & \proj{} & Ok \\ \cline{2-5}

        & \multirow{3}{27mm}{\centering androidx.biometric Library~\cite{BiometricIssue2}} &
        \multirow{3}{128mm}{AndroidX library's \texttt{BiometricManager.canAuthenticate(BIOMETRIC\_STRONG)} should return \texttt{BIOMETRIC\_SUCCESS} when fingerprint authentication is available; however, it returns \texttt{BIOMETRIC\_STATUS\_UNKNOWN} on Android 10 OS.} &
        Q (10.0) for emul. (\incompatos{}) & Error \\ \cline{4-5}
        & & & Oreo (8.1) for emul. (\compatos{}) & Ok \\ \cline{4-5}
        & & & \proj{} & Ok \\ \cline{2-5}
        
        & \multirow{3}{27mm}{\centering androidx.fragment Library~\cite{FragmentIssue}} &
        \multirow{3}{128mm}{A callback registered by \texttt{ViewCompat.setOnApplyWindowInsetsListener()} must be called once when app window is moved or resized; however, the callback is called indefinitely many times on Android 10 OS. } &
        Q (10.0) for emul. (\incompatos{}) & Error \\ \cline{4-5}
        & & & Oreo (8.1) for emul. (\compatos{}) & Ok \\ \cline{4-5}
        & & & \proj{} & Ok \\ \hline
        

        \multirow{15}{2.5mm}{\rotatebox[origin=c]{90}{Synthetic App Benchmark}} &
        \multirow{3}{27mm}{\centering Synthetic App 1} &
        \multirow{3}{128mm}{Importing \texttt{InputManager} from \texttt{input} package for using an external keyboard. \texttt{InputManager} class is moved from \texttt{inputmanager} package to \texttt{input} package with JB OS release; therefore, the app crashes with \texttt{NoClassDefFoundError} on ICS OS.} &
        ICS (4.0) for emul. (\incompatos{}) & Error \\ \cline{4-5}
        & & & JB (4.1) for emul. (\compatos{}) & Ok \\ \cline{4-5}
        & & & \proj{} & Ok \\ \cline{2-5}

        & \multirow{3}{27mm}{\centering Synthetic App 2} &
        \multirow{3}{128mm}{Using \texttt{android.app.ActivityOptions} class to set a custom screen transition animation. This causes \texttt{NoClassDefFoundError} in ICS OS because ICS OS does not have the class.} &
        ICS (4.0) for emul. (\incompatos{}) & Error \\ \cline{4-5}
        & & & JB (4.1) for emul. (\compatos{}) & Ok \\ \cline{4-5}
        & & & \proj{} & Ok \\ \cline{2-5}

        & \multirow{3}{27mm}{\centering Synthetic App 3} &
        \multirow{3}{128mm}{Calling \texttt{showChild(ViewGroup parent, View child, int oldVisibility)} to make a UI element visible. On ICS OS the app crashes with \texttt{NoSuchMethodError} since the method has a different call signature: \texttt{showChild(ViewGroup parent, View child)}.} &
        ICS (4.0) for emul. (\incompatos{}) & Error \\ \cline{4-5}
        & & & JB (4.1) for emul. (\compatos{}) & Ok \\ \cline{4-5}
        & & & \proj{} & Ok \\ \cline{2-5}

        & \multirow{3}{27mm}{\centering Synthetic App 4} &
        \multirow{3}{128mm}{Casting an \texttt{android.view.KeyCharacterMap} object into \texttt{Parcelable} Java interface to serialize it. This causes \texttt{IncompatibleClassChangeError} on ICS OS as ICS OS's \texttt{KeyCharacterMap} does not implement \texttt{Parcelable} while JB OS's does.} &
        ICS (4.0) for emul. (\incompatos{}) & Error \\ \cline{4-5}
        & & & JB (4.1) for emul. (\compatos{}) & Ok \\ \cline{4-5}
        & & & \proj{} & Ok \\ \cline{2-5}

        & \multirow{3}{27mm}{\centering Synthetic App 5} &
        \multirow{3}{128mm}{Calling \texttt{setPlaySpeed()} for media speed control when the app is on an i.MX~6 device. This causes \texttt{NoSuchMethodError} on Oreo (8.0) OS because the method is removed from KitKat (4.4) OS release without a notice.} &
        Oreo (8.0) for i.MX 6 (\incompatos{}) & Error \\ \cline{4-5}
        & & & Oreo (8.1) for emul. (\compatos{}) & Ok \\ \cline{4-5}
        & & & \proj{} & Ok \\ \hline

    \end{tabular}
    } 
    \label{tab:test_apps}
\end{table*}


We use three workload types to comprehensively evaluate whether \proj{} can recover from
compatibility crashes. First, we use the French Calendar, a 
Google Play app reported having compatibility issues by previous
studies~\cite{ficfinder, pivot, cid}.
Second, we use a prevalent Android library that we verified to
have compatibility issues, the AndroidX library. We chose this library
because (i) it is used by $84\%$ of Android apps~\cite{AndroidXStatixtics} and (ii) its
release notes are comprehensive for reproducing compatibility issues.
Lastly, we design a synthetic benchmark that consists of five apps that target
specific compatibility issues. To design this synthetic benchmark, we 
manually examined potential API mismatches between Android 4.0 and 4.1, as well
as between Android 8.1 and Android 8.0 customized for an embedded board
that we use for our evaluation, i.MX~6. We then
grouped the found mismatches into five types: missing packages, missing classes,
inconsistent method parameters, inconsistent class inheritance, and missing
methods, and developed an app exhibiting a compatibility issue for each type.

Table~\ref{tab:test_apps} summarizes our workload as well as results, which we discuss in detail below.

\noindent\textbf{Workload 1 (French Calendar):}
French Calendar is a calendar app downloaded over 10,000 times from the Google Play Store. When making a calendar event notification, it calls the \texttt{setLocalOnly()} method to avoid propagating the notification to a paired smartwatch if it exists. \texttt{setLocalOnly()} was introduced in Android KitKat (KK) (Android 4.4). French Calendar checks a user device's OS version not to call the method in OS versions below KK. However, its developer made the mistake of calling it when the OS version is lower than Android Jelly Bean (JB) (Android 4.1).
French Calendar crashes with \texttt{NoSuchMethodError} when it tries to make a calendar notification on Android Ice Cream Sandwich (ICS).
This problem lasted three months, from November 2016 to February 2017, until the app developers finally fixed it.
We have run \proj{} with JB as the compatibility OS on ICS and verified that we can successfully avoid
the crash. 


\noindent\textbf{Workload 2 (AndroidX):} 
With the AndroidX library, we reproduced two compatibility issues.
First, AndroidX's \texttt{BiometricManager} class provides \texttt{canAuthenticate(...)} method to test whether a specific authentication method is available on a device. When \texttt{BIOMETRIC\_STRONG} is passed as a parameter, the method returns \texttt{BIOMETRIC\_SUCCESS} when fingerprint authentication is available; however, it returned \texttt{BIOMETRIC\_STATUS\_UNKNOWN} on Android 10 as it sets the \texttt{mFingerprintManager} field of the \texttt{BiometricManager} class as null on Android 10 by mistake.

Second, AndroidX's \texttt{ViewCompat} class provides \texttt{setOnApplyWindowInsetsListener()} method for registering a callback to be called once when the app window is moved or resized. However, AndroidX had called the callback infinite times due to a compatibility bug when an app on Android 10 creates \texttt{FragmentContainerView} instance after calling the \texttt{WindowCompat.setDecorFitsSystemWindows()} method. 

For both cases, we ran \proj{} with Android 10 as the compatibility OS on Android 8.1 and verified
that we successfully avoided the crashes.


\noindent\textbf{Workload 3 (Synthetic Benchmark):}
We develop five apps that exhibit different compatibility issues with ICS.
App-1 uses a package that does not exist on ICS, app-2 uses a class that does not exist on ICS, app-3 calls a method with an incompatible signature, and app-4 casts a class to an incompatible subclass.
These four apps crash on ICS but not on JB.
We have run \proj{} with JB compatibility OS on ICS and verified that we can successfully
avoid the crashes.

In addition, we developed another app (app-5) that has compatibility issues with
i.MX~6's vendor-specific customization.
This customization adds the \texttt{setPlaySpeed(int)} method to the \texttt{MediaPlayer} Java class in Android OS API.
However, this method is removed in later OS releases for i.MX~6. Thus, it will crash if an app mistakenly uses
\texttt{setPalySpeed(int)} without properly handling OS versions. We have implemented
this in app-5 and verified that \proj{} can successfully avoid the crash. 

\subsection{Microbenchmark}
\label{sec:microbenchmark}

\begin{figure*}[t]
    \centering
    \begin{subfigure}[t]{0.32\linewidth}
         \centering
         \includegraphics[width=\textwidth]{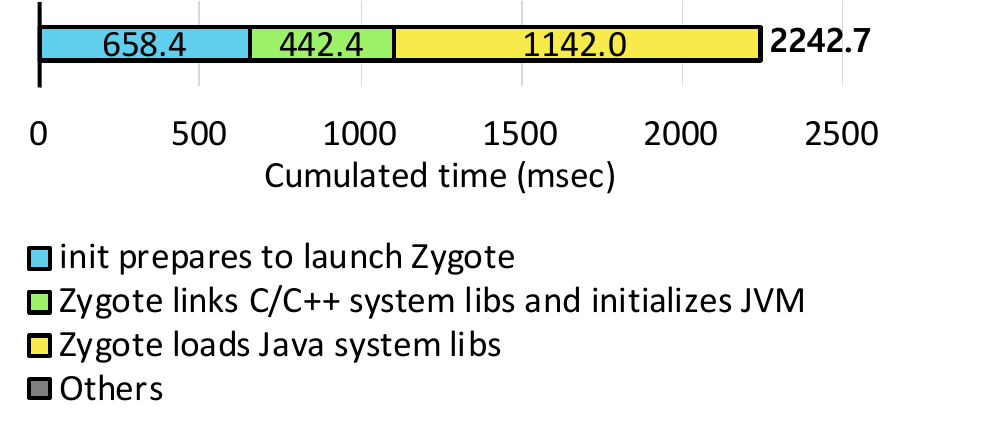}
         \Description[A horizontal bar chart breaks down the latency for booting a compatibility OS.]{A horizontal bar chart breaks down the latency for booting a compatibility OS.}
         \caption{Phase 1: Compatibility OS booting.}
         \label{fig:latency_phase1}
     \end{subfigure}
     \hfill
     \begin{subfigure}[t]{0.32\linewidth}
         \centering
         \includegraphics[width=\textwidth]{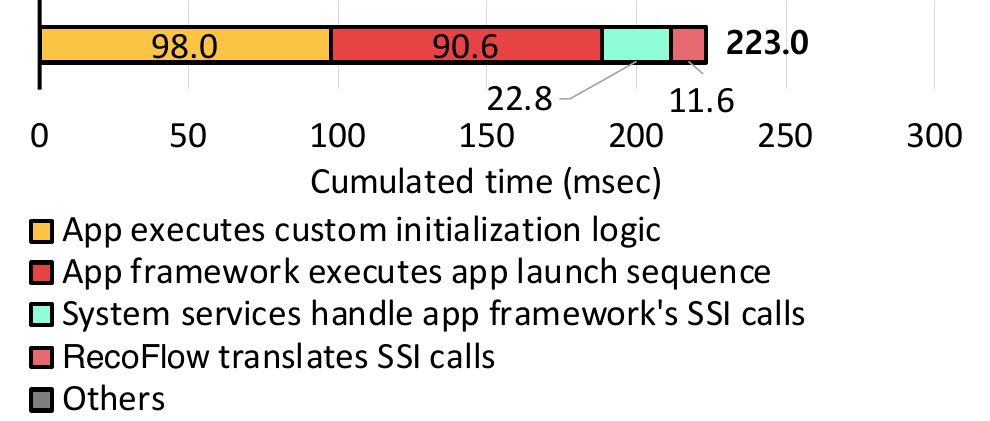}
         \Description[A horizontal bar chart breaks down the latency for launching the French Calendar app.]{A horizontal bar chart breaks down the latency for launching the French Calendar app.}
         \caption{Phase 2: FC app launch.}
         \label{fig:latency_phase2}
     \end{subfigure}
     \hfill
     \begin{subfigure}[t]{0.32\linewidth}
         \centering
         \includegraphics[width=\textwidth]{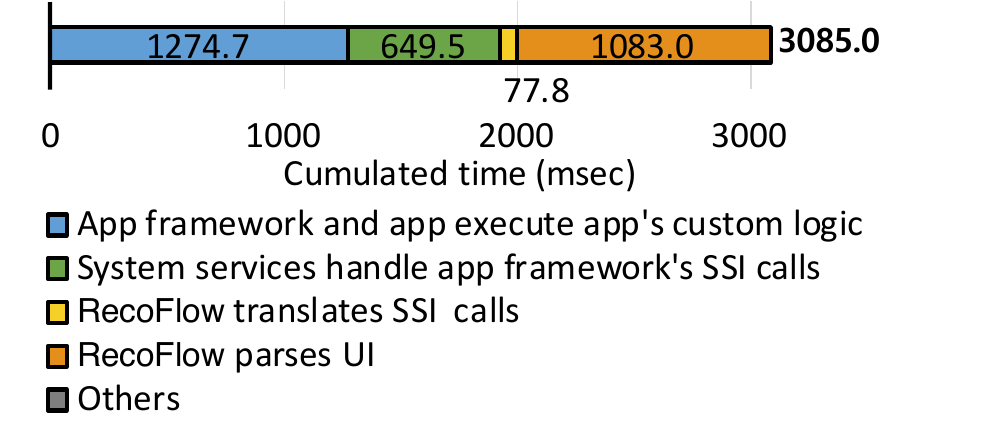}
         \Description[A horizontal bar chart breaks down the latency for recovering the French Calendar app.]{A horizontal bar chart breaks down the latency for recovering the French Calendar app.}
         \caption{Phase 3: FC app crash recovery.}
         \label{fig:latency_phase3}
     \end{subfigure}
    \caption{Latency breakdown for recovering\eat{launching} the French Calendar (FC) app's crash in compatibility mode. 
    }
    \label{fig:latency}
\end{figure*}


We evaluate \proj{}'s overhead with the French Calendar app. We modified the app's source code to use \proj{} API to recover the app's state when a crash occurs. 
As the app crashes when changing a notification setting, our user flow for the experiment has an initial stage of clicking the ``settings'' button and another stage of interacting with any UI elements in the settings window. 

We use an Android emulator running ICS for experiments. In the emulator, JB compatibility OS image and corresponding translation packs were downloaded for use. The emulator emulates a single CPU core and 501 MB RAM. The emulator is on a server with a 3.10 GHz Intel Xeon CPU.


Our experiment launched the French Calendar app on an Android emulator running ICS OS. We then click the ``settings'' to navigate to the app's settings panel and click the ``system notification'' option to change notification settings. When we click ``system notification'', the app crashes as it invokes \texttt{setLocalOnly()} Android API method that does not exist in ICS. \proj{} detects the app's crash and launches the app in compatibility mode. JB compatibility OS is booted first, and the app is then launched with the compatibility OS. As soon as the app is launched, \proj{} replays clicking the ``settings'' and ``system notification'' to restore the app's state. The app invokes \texttt{setLocalOnly()} again and successfully makes a notification as it is in compatibility mode.


\noindent\textbf{Latency overhead:} \proj{} takes 5.55~seconds 
from when the French Calendar app crashes to when the app's state is recovered in compatibility mode. We analyze the source of \proj{}'s crash handling delay in three phases; compatibility OS booting (\cref{fig:latency_phase1}), app launch (\cref{fig:latency_phase2}), and app recovery (\cref{fig:latency_phase3}) that respectively takes 2,242~ms, 223~ms, and 3,085~ms. In the compatibility OS booting phase, \texttt{init} process takes 658~ms before starting compatibility OS's Zygote. Once Zygote is started, loading C/C++ app framework libraries and initializing Java virtual machine~(JVM) take 442~ms. Loading Java app framework libraries takes 1,142~ms. In the app launch phase, the French Calendar app's custom initialization logic takes 98~ms. While the app framework and system services launch the app, they consume 91~ms and 23~ms, respectively. \proj{}'s SSI translation takes only 12~ms, which is approximately 5\% of the entire app launch phase delay. In the app recovery phase, \proj{} parses UI elements to replay recorded UI actions. 
Although this takes 1,083~ms, recovery is automatic, and the total recovery time is still short compared with a scenario that the end-user manually repeats the same UI actions to change the notification setting. System services and SSI translation take 649~ms and 78~ms, respectively. The rest of the delay, 1,273~ms, is for the app framework and the app's logic. During app launch and recovery, \proj{}'s micro-virtualization adds only 89.4~ms of SSI translation delay (2.7\%).

\begin{figure}[t]
    \includegraphics[width=\columnwidth{}]{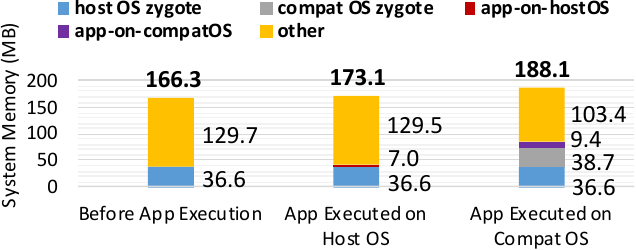}
    \Description[A vertical stack bar chart.]{A vertical stack bar chart that breaks down the memory footprints when executing the French Calendar app in the compatibility mode. The Y-axis is memory size in MB, and the x-axis is the three app execution phases, before app execution, app execution on host OS, and app execution on the compatibility OS.}
    \caption{Memory consumption breakdown.}
\label{fig:memory}
\end{figure}

\noindent\textbf{Memory \& storage overhead:} 
We measure physical memory usage of the host OS's Zygote, compatibility OS's Zygote, the French Calendar app, and any other processes. Figure~\ref{fig:memory} shows the results. Before we launch the French calendar app on the host OS, the host OS's Zygote is consuming 36.6~MB. When the app is launched, although the app uses the whole app framework that Zygote loads and initializes, the app occupies only 7.0~MB more memory. This is because the app shares memory allocated for the app framework with Zygote from when it forked from Zygote. When the app crashes and relaunches in compatibility mode, 38.7~MB additional memory is used by compatibility OS's Zygote. The compatibility mode app launches from compatibility OS's Zygote takes 9.4~MB. To summarize, \proj{} uses only 38.7~MB for compatibility OS's Zygote, which we believe is affordable for the emulator with 501~MB RAM.



As a compatibility OS image must be downloaded and stored in a device, the image adds networking and storage overhead. The JB compatibility OS image used for our evaluation is
273~MB and compresses to a 119~MB zip file.

%% file: src/10_discussion.tex
\section{Discussion}







\noindent\textbf{\proj{} on commercial devices:} 
\proj{} is deployable on commercial devices. We successfully patched vendor-customized OSes for our compatibility mode: Android ICS OS on Galaxy Nexus, Android JB OS on Galaxy Nexus, Android Lollipop OS on Galaxy S5, Android Marshmallow OS on Galaxy S5, and Android Oreo OS on Nexus 5X. We booted JB compatibility OS on ICS OS Galaxy Nexus and JB OS Galaxy Nexus. However, we could not evaluate \proj{} on these devices as
they have different layouts of partitions and file systems depending on device types (A/B device or non-A/B device) and Android OS versions~\cite{RamdiskP48:online}. Hence, their source code is required for debugging \proj{} to be fully functional.
Instead, we have evaluated \proj{} on i.MX~6, an Android development board, that the vendor provides the source code of their customized Android OS.

\noindent\textbf{Security:} One might be concerned that a malicious app could abuse \proj{}'s compatibility mode to bypass any security enforcement mechanisms implemented in the host OS. \proj{}, however, replaces only the app framework that is an app-side OS component executed as part of the app's process with the app's Unix permissions. Therefore, any attack scenarios abusing the compatibility OS's app framework is likely already possible by the app itself, and \proj{} does not add new security concerns.

%% file: src/11_conclusion.tex
\section{Conclusion}

We presented \proj{} which provides a second chance to execute a crashed app and avoid compatibility crashes in future executions. 
\proj{} enables Android app developers to automatically recover a crashed app by selectively replaying UI actions of a user intent disrupted by the crash. To do so, app developers 
use \proj{}'s easy-to-use API and \devtool{} that enables the visual programming of the user flows. Furthermore, \proj{} prevents subsequent compatibility crashes by enabling compatibility mode app execution for Android apps. In compatibility mode, Android apps are executed with a compatibility OS and avoid compatibility issues with the host OS on a device. \proj{} achieves this with its Android OS virtualization technique that effectively executes an app in a virtualized OS yet incurs only bare-minimal overhead. Our evaluation with Android app developers and Android app users demonstrates that \proj{} can effectively recover compatibility crashes under various scenarios.